\documentclass[aps,prb,amsmath,twocolumn,showpacs,superscriptaddress]{revtex4}
\usepackage{graphicx}

\usepackage{color}

\begin{document}

\title{Optical initialization, readout and dynamics of a Mn spin in a quantum dot.}

\author{C. Le Gall}
\affiliation{CEA-CNRS group "Nanophysique et semiconducteurs",
Institut N\'eel, CNRS \& Universit\'e Joseph Fourier, 25 avenue des
Martyrs, 38042 Grenoble, France}
\author{R. S. Kolodka}
\affiliation{CEA-CNRS group "Nanophysique et semiconducteurs",
Institut N\'eel, CNRS \& Universit\'e Joseph Fourier, 25 avenue des
Martyrs, 38042 Grenoble, France}
\author{C. Cao}
\affiliation{CEA-CNRS group "Nanophysique et semiconducteurs",
Institut N\'eel, CNRS \& Universit\'e Joseph Fourier, 25 avenue des
Martyrs, 38042 Grenoble, France} \affiliation{Departamento de
F\'{\i}sica Aplicada, Universidad de Alicante, San Vicente del
Raspeig, 03690 Spain}
\author{H. Boukari}
\affiliation{CEA-CNRS group "Nanophysique et semiconducteurs",
Institut N\'eel, CNRS \& Universit\'e Joseph Fourier, 25 avenue des
Martyrs, 38042 Grenoble, France}
\author{H. Mariette}
\affiliation{CEA-CNRS group "Nanophysique et
semiconducteurs", Institut N\'eel, CNRS \& Universit\'e
Joseph Fourier, 25 avenue des Martyrs, 38042 Grenoble,
France}
\author{J. Fern\'andez-Rossier}
\affiliation{Departamento de F\'{\i}sica Aplicada, Universidad de
Alicante, San Vicente del Raspeig, 03690 Spain}
\author{L. Besombes}
\email{lucien.besombes@grenoble.cnrs.fr}
\affiliation{CEA-CNRS group "Nanophysique et
semiconducteurs", Institut N\'eel, CNRS \& Universit\'e
Joseph Fourier, 25 avenue des Martyrs, 38042 Grenoble,
France}

\date{\today}

\begin{abstract}

We have investigated the spin preparation efficiency by
optical pumping of individual Mn atoms embedded in
CdTe/ZnTe quantum dots. Monitoring the time dependence of
the intensity of the fluorescence during the resonant
optical pumping process in individual quantum dots allows
to directly probe the dynamics of the initialization of the
Mn spin. This technique presents the convenience of
including preparation and read-out of the Mn spin in the
same step. Our measurements demonstrate that Mn spin
initialization, at zero magnetic field, can reach an
efficiency of $75\%$ and occurs in the tens of \emph{ns}
range when a laser resonantly drives at saturation one of
the quantum dot transition. We observe that the efficiency
of optical pumping changes from dot to dot and is affected
by a magnetic field of a few tens of mT applied in Voigt or
Faraday configuration. This is attributed to the local
strain distribution at the Mn location which predominantly
determines the dynamics of the Mn spin under weak magnetic
field. The spectral distribution of the spin-flip scattered
photons from quantum dots presenting a weak optical pumping
efficiency reveals a significant spin relaxation for the
exciton split in the exchange field of the Mn spin.

\end{abstract}

\pacs{78.67.Hc, 78.55.Et, 75.75.+a}

\maketitle

\section{Introduction.}

The ability to control spins in semiconductors
nanostructures is an important issue for spintronics and
quantum information processing. Single spin detection and
control is a key but very challenging step for any
spin-based solid-state quantum computing device. In the
last few years, efficient optical techniques have been
developed to control the spin of individual carriers
\cite{Press2008} or ensemble of nuclei \cite{Latta2009} in
semiconductor quantum dots (QDs). Thanks to their expected
long spin coherence time, magnetic atoms in a semiconductor
host could be an alternative media to store quantum
information in the solid state. However, as these localized
spins interact weakly with their environment, they can be
hardly controlled by electrical or optical methods.
Recently QDs containing individual Mn atoms have been
realized both in II-VI \cite{Besombes2004} and III-V
\cite{Kudelski2007} compounds. In these systems, since the
confined carriers and Mn spin functions become strongly
mixed, the resonant optical excitation of the QD strongly
affects the spin state of Mn atom offering a possibility of
full optical control \cite{Reiter2009}.

When a Mn atom is included in a II-VI semiconductor QD
(CdTe in ZnTe), the spin of the optically created
electron-hole pair (exciton) interacts with the five {\it
d} electrons of the Mn (total spin S=5/2). This leads to a
splitting of the once simple photoluminescence (PL)
spectrum of an individual QD into six (2S+1) components.
This splitting results from the spin structure of the
confined heavy holes which are quantized along the QDs'
growth axis with their spin component taking only the
values J$_z$=$\pm$3/2. In first approximation, the hole-Mn
exchange interaction reduces to an Ising term J$_z$.S$_z$
and shifts the emission energy of the QD, depending on the
relative projection of the Mn and hole spins
\cite{Leger2007}. As the spin state of the Mn atom
fluctuates during the optical measurements, the six lines
are observed simultaneously in time average PL spectra. The
intensities of the lines reflect the probability for the Mn
to be in one of its six spin components and the PL is a
probe of the spin state of the Mn when the exciton
recombines \cite{Besombes2008}.

In this article, we show that one can exploit the
absorption of an individual II-VI QD to optically
initialize the spin state of an embedded Mn atom. We use
resonant optical excitation of one of the six exciton
levels of a Mn-doped QD to prepare by optical pumping the
spin state of the magnetic atom. Under these resonant
excitation conditions, scattered photons coming from
spin-flips of the exciton without change of the Mn spin are
observed. This fluorescence signal is used to probe the
dynamics of the initialization of the Mn spin during the
optical pumping process. A pumping efficiency of about 75\%
is obtained with an initialization time in the tens of
\emph{ns} range. A saturation of the optical pumping
efficiency is observed at high excitation power reflecting
the non-linear absorption of the QD exciton state. We
observed that the efficiency of optical pumping is affected
by a magnetic field of a few tens of mT applied in Voigt or
Faraday configuration, and that it can be very weak at zero
field for some of the QDs. This is attributed to the local
strain distribution at the Mn location which determines
predominantly the dynamics of the Mn spin. The spectral
distribution of the photons scattered by the QD shows that
an efficient exciton spin relaxation occurs within the
exciton-Mn complex. This spin relaxation is attributed to
the large single-phonon mediated spin-flip for an exciton
split in the exchange field of the Mn.

The manuscript is divided in six sections: In section II we
give details about the Mn-doped QD samples and the time
resolved optical pumping experiments. In section III we
describe how angular momentum is transferred from
circularly polarized light to a Mn spin in a QD. In section
IV and V we show how the time resolved and cw resonant
fluorescence signal can be used to probe the dynamics of
the exciton and Mn spins. In section VI we present a simple
model for the optical pumping process and discuss the
mechanism controlling the Mn spin dynamics.

\section{Samples and experiment.}

The samples used in this study are all grown on ZnTe
substrates
 and contain CdTe/Zn$_{0.8}$Mg$_{0.2}$Te QDs. A 7.5 monolayer thick
  CdTe layer is deposited at 280$^{\circ}$C by atomic layer epitaxy
   on a Zn$_{0.8}$Mg$_{0.2}$Te barrier grown by molecular beam epitaxy
    at 360$^{\circ}$C. The dots are formed by the well established Tellurium
     deposition/desorption \cite{Tinjod2003} process and protected by a 100nm
       thick Zn$_{0.8}$Mg$_{0.2}$Te top barrier. The incorporation of Mg in the
        barriers leads to a stronger confinement of the holes and consequently to
         an increase of their exchange interaction with the Mn spin. The height of
          the QDs' core is of few nm and their diameter is in the 10-20 nm range. Mn
           atoms are introduced during the growth of the CdTe layer. The Mn concentration
            is adjusted to optimize the probability to detect one Mn per dot \cite{Maingault2006}.

During this growth process, the position of the magnetic
atom, the shape of the QD and the strain at the Mn location
are not controlled. In most of the dots, the presence of an
anisotropic strain distribution in the QD plane produces a
mixing of light-hole (lh) and heavy-hole (hh) subbands, as
described with  the Bir-Pikus Hamiltonian. This
valence-band mixing (VBM) is responsible for a linear
polarization degree of the QD emission with a linear
polarization direction imposed by the strain distribution
\cite{Koudinov2004}. In QDs containing a magnetic atom, the
VBM affects the coupling between the confined hole and the
localized magnetic moments. It allows simultaneous hole-Mn
spin flips responsible for a coupling between the dark and
bright exciton states at zero magnetic field. This is
responsible for the appearance of additional PL lines (more
than 6) on the low energy side of the emission spectrum of
most of the Mn doped QDs \cite{Leger2007} as it will be
detailed in section V.

Optical addressing of QDs containing a single magnetic atom
are achieved using micro-spectroscopy techniques. High
refractive index hemispherical solid immersion lens are
mounted on the bare surface of the sample to enhance the
spatial resolution and the collection efficiency of single
dot emission in a low-temperature ($T$=5K) scanning optical
microscope. This technique reduces the reflected and
scattered light at the sample surface and decrease the
focal spot area allowing the measurement of spin-flip
scattered photons from an individual Mn-doped QD. Despite
the quite large QD density and the large number of dots in
the focal spot area, single QD transitions can be
identified by their spectral signatures (see inset of
Fig.~2). A weak magnetic field (a few tens of $mT$) can be
applied in Voigt or Faraday configuration using either
permanent magnets or superconductive coils.

In the time resolved optical pumping experiments presented
in this article, single QD PL is quasi-resonantly
(\emph{probe}) and resonantly (\emph{pump}) excited with
two tunable continuous wave (\emph{cw}) dye lasers. Trains
of resonant light pulses with variable duration and
wavelength are generated from the \emph{cw} lasers using
acousto-optical modulators with a switching time of about
10 $ns$. The circularly polarized collected PL is dispersed
by a 1 $m$ double monochromator before being detected by a
fast avalanche photodiode in conjunction with a time
correlated photon counting unit with an overall time
resolution of about 50ps.

\section{Resonant optical pumping of a single Mn spin.}

\begin{figure}[hbt]
\includegraphics[width=3.1in]{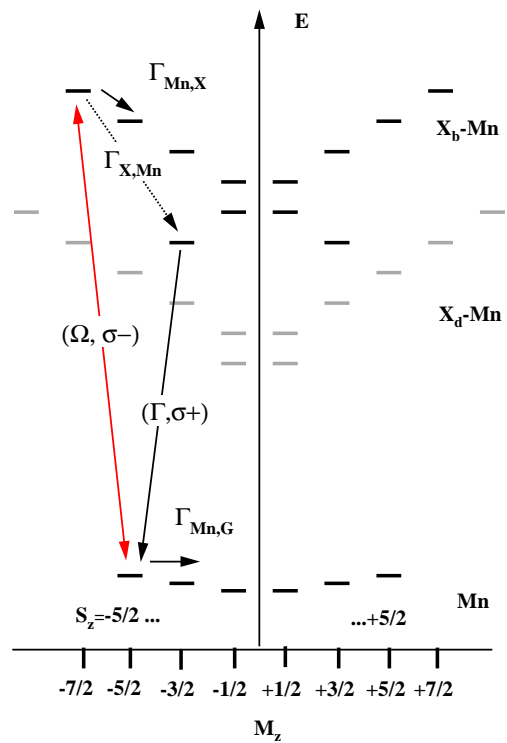}
\caption{Energy levels of a Mn doped QD involved in the
optical pumping mechanism described in the text (Black:
bright excitons (X$_b$)); grey: dark excitons (X$_d$)). The
sates are displayed as a function of their total angular
momentum M$_z$ and energy E. For the resonant optical
pumping, the QD is resonantly driven on the state
$S_z=-5/2$ with a $\sigma-$ laser pulse with a Rabi
frequency $\Omega$. The scattered photons obtained after a
spin-flip of the exciton (rate $\Gamma_{X,Mn}$) are
recorded in $\sigma+$ polarisation. The intensity of the
$\sigma+$ PL detected on the low energy bright exciton
level is proportional to the population of $S_z=-5/2$ and
is used to probe the optical pumping efficiency.}
\label{fig1}
\end{figure}

Recently, the optical orientation of a single magnetic atom
by the injection of spin polarized carriers in a Mn-doped
II-VI QD has been demonstrated \cite{LeGall2009}. In this
experiment, spin polarized excitons were generated on a QD
excited state and, after relaxation, a non-equilibrium Mn
spin population was produced by a thermalization among the
levels of the exciton-Mn (X-Mn) ground states. However, as
proposed by Govorov {\it et al.} \cite{Govorov2005}, the
direct resonant excitation of one optical transition of the
ground X-Mn complex could be used to perform a direct and
more efficient optical pumping of the Mn spin. In this
optical pumping process, a laser drives resonantly one of
the exciton-Mn transition ( $|-1,S_z=-5/2\rangle$ in
Fig.~1) with a Raby frequency
$\Omega=\mathcal{P}\mathcal{E}/\hbar$ with $\mathcal{P}$
the dipolar moment of the QD transition and $\mathcal{E}$
the amplitude of the electric field of the resonant laser.
A photon absorption occurs only if the Mn spin in the QD is
in the $S_z=-5/2$ spin state. The resultant exciton can
radiatively recombine via the same channel, or a relaxation
process can project the X-Mn complex in a state with
$S_z\neq-5/2$.
   After a few cycles of absorption-emission, the probability
   of detecting the Mn in the $S_z=-5/2$ state decreases. In this mechanism, we have
    assumed that the Mn spin was conserved once the exciton has recombined.
    The conservation of the Mn spin between the recombination of an exciton and
    the absorption of a photon can be altered in two ways: either by a relaxation
    process involving an exchange of energy or by a coherent evolution \cite{LeGall2009,Goryca2009}.
     A coherent evolution can be neglected if the fine
structure of the Mn atom is dominated
     by a magnetic anisotropy along the growth axis \cite{LeGall2009}. Otherwise,
     processes such as the coherent evolution of the Mn spin in the hyperfine field
      of the Mn nucleus or in the tetragonal crystal field leads to a change of the Mn
       spin state between the injection of two excitons. In that case, no optical
       pumping can occur. In the fallowing, we will use $\Gamma_{Mn,G}$ to describe the
       characteristic rate at which the Mn spin state changes due to coherent or incoherent
       processes, when the QD is empty. $\Gamma_{Mn,X}$ will
       correspond to the relaxation rate of the Mn interacting with an exciton. This mechanism of Mn spin manipulation is similar to the
pumping process used to prepare a single carrier spin in a
QD \cite{Dreiser2008, Gerardot2008}. It involves a
forbidden transition (\emph{i.e.} spin flip of the Mn
interacting with the exciton) and is based on the
inequality $\Gamma_{Mn,X} > \Gamma_{Mn,G}$.

To demonstrate and test the efficiency of this optical
pumping process, we developed a two wavelength pump-probe
set-up allowing an optical initialization and read-out of
the Mn spin. In this experiment, a resonant circularly
polarized $cw$ laser (\emph{resonant pump}) tuned on a X-Mn
level pumps the Mn spin. In the initial state at t=0, the
Mn atom is in thermal equilibrium (\textit{i.e.} all the
six spin states are equally populated). The resonant
creation of an exciton followed by a spin relaxation of the
Mn in the exchange field of the exciton empties the
spin-state under excitation: the probability of occupation
of this spin state decreases over time. Then, a second
laser train, linearly polarized and tuned on an excited
state of the QD (\emph{quasi-resonant probe}), injects
excitons independently of the Mn spin state $S_z$. Spin
relaxation of the X-Mn complex under these conditions of
excitation drives the Mn atom back to an equilibrium where
all spin states are equally populated. Recording one of the
six PL lines under this periodic sequence of excitation, we
monitor the time evolution of the probability of occupation
of a given Mn spin state.

\begin{figure}[hbt]
\includegraphics[width=3.2in]{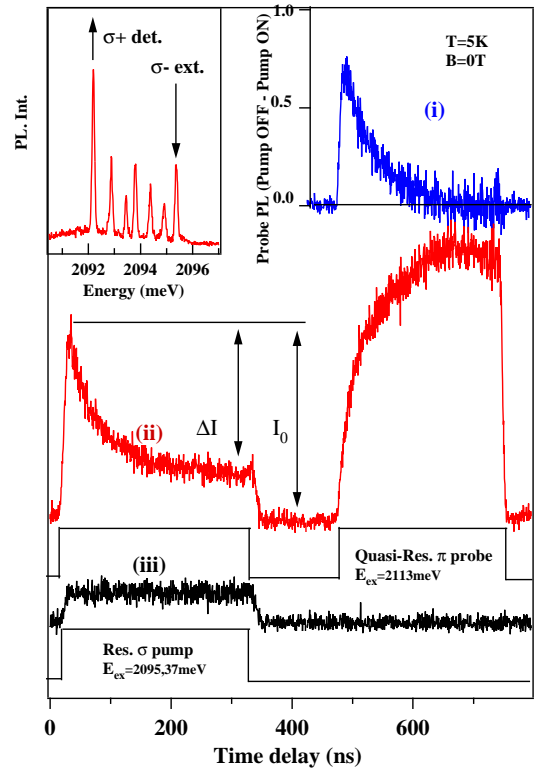}
\caption{ (Color online) PL transients recorded on the low
energy line of a Mn-doped QD (QD1) under the quasi-resonant
(QD excited state: $probe$) and resonant  (QD ground state:
$pump$) optical excitation sequence displayed at the
bottom. The inset presents the QD PL under non-resonant
excitation and the configuration of the resonant excitation
and detection. (i) Difference between the PL produced by
the probe when the pump is OFF and when the pump is ON,
(ii) PL from the pump and the probe and (iii) resonant PL
produced by the pump alone. Because of the Mn spin memory
in the absence of injected carriers, no signature of
pumping is observed when the linearly non-resonant probe is
OFF (iii). The optical pumping process is directly observed
on the resonant fluorescence produced by the pump and
latter on the PL from the probe laser. I$_0$ is the
amplitude of the fluorescence at the beginning of the pump
pulse and $\Delta I$ the amplitude of the transient.
$\Delta I /I_0$ is the efficiency of the spin optical
pumping.} \label{fig1}
\end{figure}

The main features of this experiment are presented in
Fig.~2. In this example, $\sigma+$ PL signal is recorded on
the low energy X-Mn line. The QD is resonantly excited on
the high energy state of the X-Mn complex with $\sigma-$
photons. This excitation can only create an exciton in the
dot if the Mn spin state is $S_z$=-5/2. This selective
absorption combined with a spin relaxation for the X-Mn
complex are responsible for the optical pumping process
that progressively decreases the population of the state
$S_z$=-5/2. After this pumping sequence, the resonant pump
laser is switched off and followed by a linearly polarized
excitation on an excited state (\emph{quasi-resonant
probe}). The amplitude of this quasi-resonant PL depends on
the population of $S_z$=-5/2 and, at the beginning of the
probe pulse, is a probe of the resonant pumping efficiency
reached at the end of the pump pulse.

This is illustrated in Fig.~2(i) which presents the
difference of the two PL signals produced by the probe when
the resonant pump laser was OFF or ON in the pump-probe
sequence presented underneath the curve 2(ii). The
difference of the two PL signals reflects the population
difference between a sequence with optical pumping and a
sequence where $S_z=-5/2$ is evenly populated. The height
of the difference signal at the beginning of the probe
pulse, which reaches $75\%$ gives a direct measurement of
the efficiency of the spin optical pumping. The PL
transients observed during the probe pulse corresponds to
the progressive destruction of the non-equilibrium
distribution prepared by the pump. This reset process is
produced by the injection of unpolarized excitons and its
rapidity depends on the intensity of the probe laser.

A more direct way to probe the optical pumping process is
to monitor the time evolution of the fluorescence signal
observed during the resonant excitation. Excitation
transfer can occur within the X-Mn complex during the
lifetime of the exciton and gives rise to a weak PL on all
the QDs levels. Whatever the spin relaxation processes
involved in this excitation transfer, this signal depends
on the absorption of the pump laser which is controlled by
the occupation of $S_z$=-5/2: it monitors the spin
selective absorption of the QD and is then a direct probe
of the pumping efficiency of the Mn spin. The pumping
efficiency is then given by $ \Delta I / I_0\approx75\% $
(see Fig.~2), in agreement with the pumping efficiency
measured on the probe sequence.

The time evolution of the PL detected on the low energy
state of X-Mn under a resonant excitation on the high
energy state is presented in Fig.~2(ii) and 2(iii) for two
different pump-probe sequences: probe ON and probe OFF
respectively. When the probe laser is switched ON in the
pump-probe sequence, an equilibrium distribution of the Mn
spin is restored by the non-resonantly injected unpolarized
excitons before each pumping pulse. The absorption, and
then the amplitude of the resonant fluorescence signal is
maximum at the beginning of the pump pulse and
progressively decreases as the state $S_z$=-5/2 is emptied
by the optical pumping process. When the probe laser is
switched OFF in the pump-probe sequence, the resonant
fluorescence transients during the pump pulse disappears
and a weak constant PL is observed. This disappearance of
the transient is a signature of the perfect conservation of
the Mn spin distribution during the dark time between each
pumping pulse. The steady state PL depends on the optical
pumping efficiency which is controlled by the ratio of the
relaxation rates for the Mn spin in the exchange field of
the exciton and the relaxation and coherent evolution of
the Mn spin in an empty dot \cite{LeGall2009,Goryca2009}.

\begin{figure}[hbt]
\includegraphics[width=3.4in]{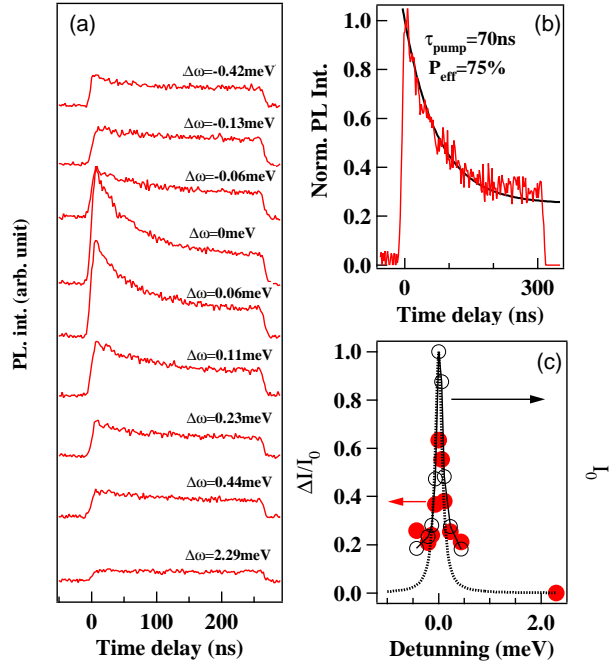}
\caption{(color online) (a) Excitation energy dependence of
the resonant fluorescence signal obtained for cross
circular excitation-detection on the high and low energy
line respectively (positive detuning corresponds to an
excitation on the hight energy side of the line). (b)
Detail of the resonant fluorescence transient recorded
during the optical pumping process. The exponential fit
(black line) gives an optical pumping efficiency
$P_{eff}$$\approx$75$\%$ and a pump time of 70 ns. (c)
Amplitude of the resonant fluorescence signal as the
excitation is tuned around the high energy line of X-Mn.
The Lorentzian fit give a full width at half maximum of
80$\mu eV$.} \label{fig2}
\end{figure}

\section{Time resolved resonant fluorescence.}

The resonant fluorescence signal can be used to analyse the
influence of the excitation intensity, wavelength and
polarisation on the efficiency of the Mn spin optical
pumping. A detail of the time resolved resonant
fluorescence signal obtained with the pump laser tuned
strictly on resonance with the high energy level is
presented in Fig.~3(b). A decrease of about 75\% of the
resonant PL is observed during the optical pumping process
with a characteristic time of $\tau_{pump}$=70 $ns$. This
exponential decay reflects the decrease of the absorption
of the QD induced by the decrease of the population of the
state $S_z$=-5/2 and shows it takes a few tens of $ns$ to
initialize the Mn spin. Alternatively, one can say that the
transition can be recycled for a few tens of $ns$ before
the laser induces a Mn spin-flip event. After a few tens of
$ns$ the PL reaches a steady state intensity. This resonant
fluorescence signal can be used to analyze in detail the
optical pumping mechanism.

Fig.~3(a) and 3(b) present the amplitude and time evolution
of the fluorescence signal detected on
$|+1,S_z=-5/2\rangle$ for different $pump$ wavelength
around the high energy level $|-1,S_z=-5/2\rangle$. A clear
resonant behavior is observed in the initial amplitude
$I_0$ of the fluorescence signal (Fig.~3(c)). This reflects
the wavelength and excitation power dependence of the
absorption of the QD. The measured width of the resonance
($\sim 80\mu eV$) is a convolution of the width of the QD's
absorption in the non-linear regime and of the linewidth of
the excitation laser. The efficiency of the optical pumping
$\Delta I/I_0$, presents a similar resonance demonstrating
the strong excitation energy dependence of the spin optical
pumping process.

\begin{figure}[hbt]
\includegraphics[width=3.5in]{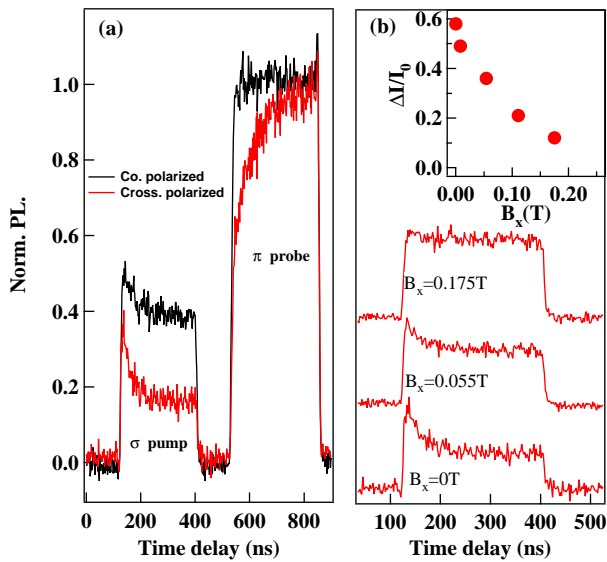}
\caption{(color online) (a) Circularly polarized PL obtained during
the optical pumping sequence for co and cross circularly polarized
pump. (b) Transverse magnetic field (B$_x$) dependence of the
resonant fluorescence signal under cross polarized
excitation-detection. The inset presents the amplitude of the
optical pumping signal as a function of transverse magnetic
field.}\label{fig3}
\end{figure}

Fig.~4(a) presents the time resolved PL recorded on the low
energy line during the pump and the probe pulses detected
in the two circular polarizations. For cross circularly
polarized pump excitation and PL detection, the PL probes
the Mn spin state resonantly excited by the pump laser
($S_z$=-5/2 for $\sigma$- pump). In this case, the pumping
effect is both observed in the transient of the resonant
fluorescence and as a decrease of the initial PL intensity
in the probe signal. For co-polarized pump excitation and
PL detection ($\sigma-$), the PL intensity of the low
energy line is proportional to the population of
$S_z$=+5/2. As evidenced by the PL of the probe pulse, the
population of this state is not significantly affected by
the resonant pump laser. The resonant PL during the pump is
dominated by the direct absorption in the low energy line
acoustic phonon side-band \cite{Besombes2001} and the
transient caused by the depletion of $S_z$=-5/2 can be
hardly observed. As expected for a Mn pumping process, the
influence of a $\sigma$- resonant laser tuned on the high
energy X-Mn level mainly affects the population of the
state $S_z$=-5/2.

\begin{figure}[hbt]
\includegraphics[width=3.4in]{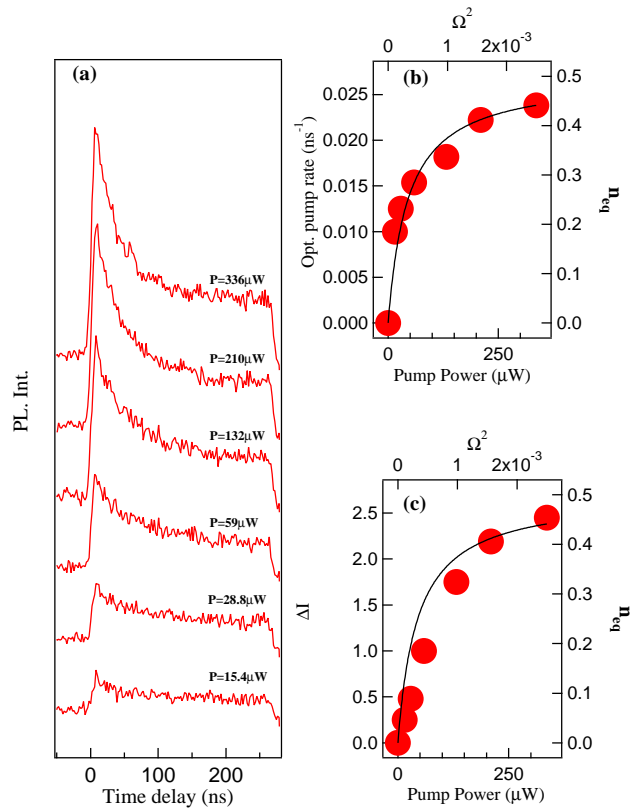}
\caption{(color online)(a) Excitation power dependence of the
resonant fluorescence signal. (b) Excitation power dependence of the
optical pumping rate. (c)  Excitation power dependence of the
amplitude of the optical pumping signal. The solid line describe the
population saturation of a resonantly excited two level system.}
\label{fig4}
\end{figure}

More information on the pumping process can be obtained
from the transverse magnetic field dependence of the time
resolved resonant fluorescence signal. The precession of
the Mn spin in a transverse magnetic field decreases the
probability of conserving a prepared $S_z$ state. For an
isotropic Mn spin, the damping of the precession induced by
the decoherence should give rise to the standard Hanle
depolarization curve with a Lorentzian shape and a width
proportional to $1/T_2$ \cite{Myers2008} where $T_2$ is the
decoherence time of Mn spin. In the present case, an
in-plane magnetic field $B_x$ induces coherent precession
of the Mn spin away from the optical axis (=~QDs' growth
axis), so that the average spin polarization, and therefore
the amplitude of the optical pumping signal, decays
(Fig.~4(b)). However, a strain-induced magnetic anisotropy
along the growth axis \cite{LeGall2009,Qazzaz1995}
partially blocks the Mn spin precession. As already
observed in the optical orientation experiments under
non-resonant excitation (see Ref. 15), the transverse
magnetic field decreases the efficiency of the optical
pumping (inset of Fig.~4(b)). This field dependence is not
controlled by the Mn coherence time but mainly by its
magnetic anisotropy and is a probe of the strain state of
the QD at the Mn location.

As displayed in Fig.~5, the characteristic time and the
amplitude of the optical pumping signal also depends on the
excitation intensity. In the low excitation regime, as
expected for a spin optical pumping process, the transient
characteristic time ($\tau_{pump}$) is inversely
proportional to the pump laser intensity. However, a
saturation behavior is clearly observed both for the
amplitude and the characteristic time of the resonant
fluorescence transient. Here, the rate of the spin optical
pumping saturates around 0.025 $ns^{-1}$ (Fig.~5(b)). The
saturation of the optical pumping process results from a
saturation of the absorption of the resonantly excited
excitonic level. Indeed, the population of a two level
system driven by a resonant excitation laser is given by:

\begin{equation}
n_{eq}=\frac{1}{2}\frac{\Omega^2(\frac{T_1}{T_2})}{(\Delta\omega^2+\frac{1}{T_1^2}+\Omega^2\frac{T_1}{T_2})}
\label{raby}
\end{equation}

\noindent where $\Omega$ is the Rabi frequency,
$\Delta\omega$ the detuning between the excitation laser
and the excitonic transition, T$_{1}$ and T$_{2}$ the
lifetime and the coherence time of the exciton
respectively. The rate of the spin optical pumping process,
which is proportional to $n_{eq}$, is expected to increase
with the excitation Rabi frequency until it reaches a
saturation value when the Rabi frequency is larger than the
spontaneous emission rate ($\Omega$ $\gg$ $T_1^{-1}$). The
saturation curve obtained with equation (\ref{raby}),
T$_{1}$=180ps, T$_{2}$=10ps \cite{Patton2006} and
$\Delta\omega$=0 is compared with the optical pumping
signal in Fig.~5(b) and 5(c). A good agreement with this
simple model describing the population of a two level
system resonantly excited by a \emph{cw} laser is obtained.

In the saturation regime, if $S_z=-5/2$ the QD is in the
$|-1,S_z=-5/2\rangle$ state half of the time in average.
Taking for granted that $\Gamma_{Mn,X} \gg \Gamma_{Mn,G}$,
the rapidity of the optical pumping process is no longer
controlled by the rate at which excitons are injected but
depends only on the relaxation rate from the state
$|-1,S_z=-5/2\rangle$ to other X-Mn levels with
$S_z\neq-5/2$. Therefore, the pumping rate in the
saturation regime gives an estimation of the spin-flip rate
of the Mn in the exchange field of the exciton
$\Gamma_{Mn,X}\approx \Gamma_{pump}/2$ and a relaxation
time $\tau_{Mn,X}\approx 80 ns$.

\section{Dynamics of exciton and Mn spins.}

Information about the spin relaxation mechanism of the
exciton exchange coupled with a Mn spin can be obtained
from the energy of the resonant fluorescence signal. The
spectral distribution of the scattered photons observed
during the resonant excitation on the QD ground state
results from spin-flip processes within the exciton-Mn
systems. A simple thermalization among the 24 exciton-Mn
levels (see ref. 11) should give rise to a thermal
distribution of the intensities of the PL lines in the
resonant fluorescence spectrum.

Here, we report results on a QD presenting a weak optical
pumping at $B=0T$. This poor efficiency of optical pumping
could be attributed to the local strain environment which
leads to $\Gamma_{Mn,X}\approx\Gamma_{Mn,G}$. This case is
interesting for a study of the $X-Mn$ dynamics: The QD line
under excitation is always absorbant and the scattered
photons reflect the fastest spin relaxation channels.
Fig.~6(a) presents the three photoluminescence excitation
spectra (PLE) detected on the low energy lines of X-Mn
which are associated with the Mn spin states +5/2, +3/2 and
+1/2 (plain lines). In addition, PLE spectra detected on a
dark exciton state associated with the Mn spin $\pm$1/2 is
also presented (dotted line). The corresponding resonant PL
spectra obtained at the maximum of the PLE signal are also
displayed. Finally, a PL spectrum in quasi-resonant
excitation is displayed at the bottom of Fig.~6(a). The
QD's spectrum differs from the one expected in the heavy
hole approximation and presents seven lines. Such PL
spectrum appears when the high energy lines of the dark
exciton states overlap the low energy lines of the bright
exciton states. Strained induced VBM allows simultaneous
hole-Mn spin-flips and couples dark and bright states. If
the states coupled through these spin-flips are close in
energy in the heavy hole approximation, this leads to a
strong mixing of a bright state with a dark state and the
QD presents additional PL lines. This is the case for the
QD presented in Fig.~6(a): VBM mixes $|+1,-3/2\rangle$ with
$|-2,-1/2\rangle$ and the new eigenstates share the
oscillator strenght of the bright state $|+1,-3/2\rangle$.
The two lines on the right of the low energy state can be
attributed to the bright part of the states
$\alpha|+1,-3/2\rangle+\beta|-2,-1/2\rangle$ and
$\beta|+1,-3/2\rangle+\alpha|-2,-1/2\rangle$, with
$|\alpha|$ and $|\beta|$ close to $1/2$. The two small
peaks at low energy come from dark states with a weaker
bright exciton component. The attribution of these lines to
the bright and dark exciton states is confirmed by the
calculation of the energy levels presented in Fig.~6(b)
following ref. 9 and 24.

From these PLE spectra and resonant PL spectra, it follows
that the most efficient spin relaxation channels within the
X-Mn system do conserve the Mn spin. For instance, an
excitation with $\sigma+$ photons on the high energy line
(state $|+1,S_z=+5/2\rangle$) produces $\sigma-$ PL mainly
on the low energy state $|-1,S_z=+5/2\rangle$. A similar
behavior is observed for an excitation on
$|+1,S_z=+3/2\rangle$ which generate a PL on
$|-1,S_z=+3/2\rangle$. In both cases, this excitation
transfer corresponds to a spin flip of the exciton from
$|+1\rangle$ to $|-1\rangle$ with a conservation of the Mn
spin.

\begin{figure}[h]
\includegraphics[width=3.7in]{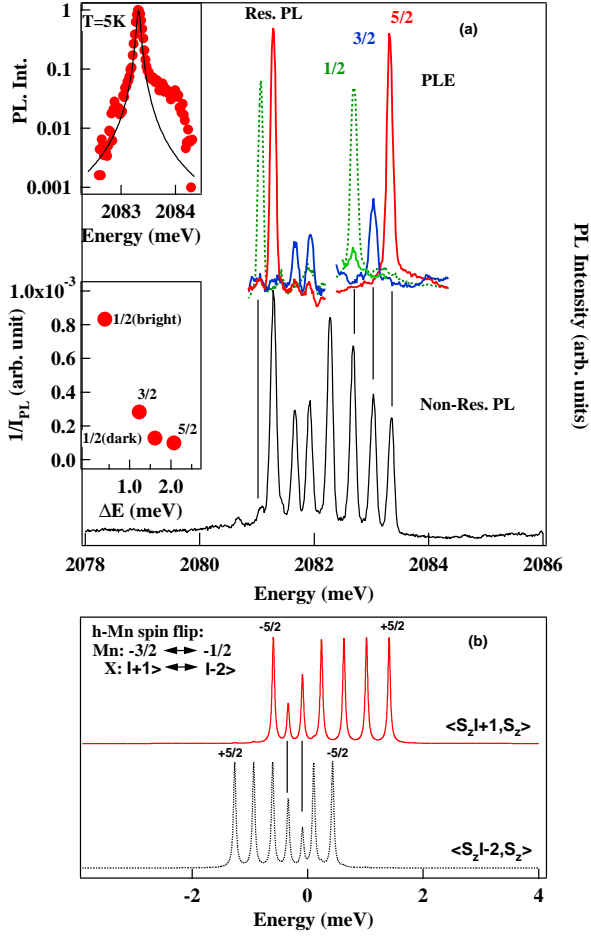}
\caption{(color online) Photoluminescence excitation (PLE)
spectra and resonant photoluminescence obtained on the
ground state of a Mn doped QD (QD2). The resonant PL is
obtained under circularly cross-polarized
excitation-detection. The spin state of the Mn is conserved
during the spin flip of the exciton. The inset presents an
enlarge view of the PLE obtained under excitation and
detection on the $S_z$=+5/2 state. The black line is a
Lorentzian fit with a half width at half maximum of
80$\mu$eV. The asymmetry of the absorption line comes from
a coupling with acoustic phonons. (b) Calculated bright
$|+1\rangle$ and dark $|-2\rangle$ energy levels with
$J_{eMn}$=-0.06 meV, $J_{hMn}$=0.245 meV ,$J_{eh}$=-0,550
meV and $\epsilon_{vbm}=0.1$. Because of the large
carrier-Mn exchange coupling, the dark exciton levels
overlap with the bright excitons. The valence band mixing
coupled the states $|+1,-3/2\rangle$ and $|-2,-1/2\rangle$
and gives rise to an additional PL line.}\label{fig6}
\end{figure}

More surprising, an excitation on $|+1,S_z=+1/2\rangle$
produces the strongest PL on an exciton level formed mostly
of a dark state with a weak oscillator strength (weak
signal in the non-resonant PL plotted at the bottom of
Fig.~6(a)) with the same Mn spin ($S_z$=+1/2). This
confirms the good stability of the Mn spin under these
resonant excitation conditions and the efficient spin-flip
of the exciton in the exchange field of the Mn.

In the spin-flip scattering processes which conserve the Mn
spin, the \emph{hh} exciton stays in the same spatial state
and just flips its spin ({\it i.e.} the electron and the
hole spin flip simultaneously), by simultaneously emitting
(or, at high temperature, absorbing) an acoustic phonon.
This can occur via the \emph{lh} and \emph{hh} exciton
mixing due to the interplay of the short range exchange
interaction and the lattice deformation \cite{Leger2007}.
Such single phonon process is responsible for an increase
of the exciton spin relaxation rate observed in small QDs
with the increase of the Zeeman splitting $\Delta E$
\cite{Tsitsihvili2003,khaetskii2001,woods2002}.

\begin{figure}[hbt]
\includegraphics[width=3.3in]{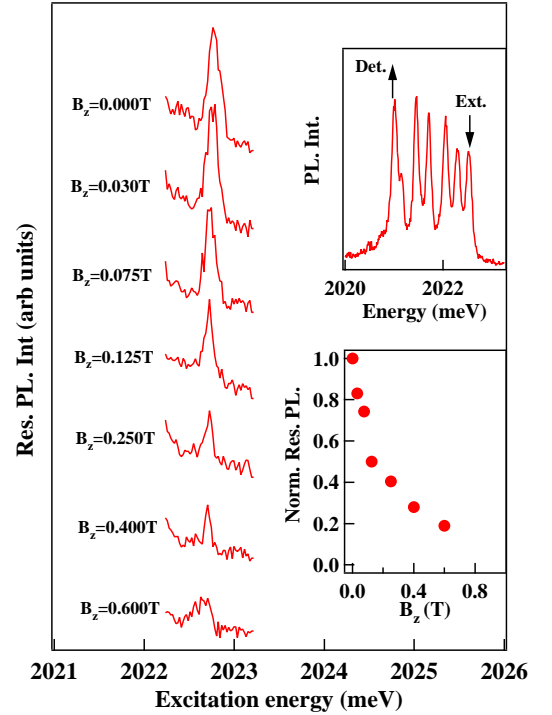}
\caption{(color online) Magnetic field dependence of the
photoluminescence excitation (PLE) spectra obtained on the
ground state of a Mn doped QD (QD3) under circularly
cross-polarized excitation-detection. The insets present a
PL spectra of QD3 (arrows point the excitation and
detection wavelength) and the magnetic field ($B_z$)
dependence of the amplitude of the resonant PL signal.}
\label{fig6}
\end{figure}

For typical QDs and low temperature, the zero magnetic
field exciton spin relaxation time is longer than the
exciton radiative lifetime and no exciton spin-flip is
expected during its lifetime. However for QDs with large
short-range exchange interaction and small \emph{hh-lh}
splitting like CdTe/ZnTe QDs \cite{Leger2007}, an increase
of the exciton splitting can significantly shorten this
relaxation time \cite{Tsitsihvili2003,woods2002}. This
faster spin relaxation rate is a consequence of the
increase of the acoustic phonon density of states at the
energy of the inter-level splitting.

In Mn-doped QDs, the exciton Mn exchange interaction acts
on the exciton as an effective magnetic field, aligned
along the QDs' growth axis, which increases with $S_z$.
With values of $\Delta_{st}\sim$ 1 meV, $\Delta_{lh-hh}\sim
30 meV$ \cite{Leger2007} and an inter-level splitting
$\Delta E=2meV$ for the $S_z=\pm5/2$ states in QD2, we
estimate according to reference 20, an exciton spin
lifetime of about 0.4 ns. The large exciton splitting leads
to an exciton spin lifetime that can reach the exciton
lifetime scale $\approx$0.2ns. This decrease of the exciton
spin lifetime, which scales as $(\Delta E)^{-3}$ (according
to ref. 19), can explain the observed efficient transfer of
excitation within the X-Mn complex. This splitting
dependence of the exciton relaxation rates can also explain
the decrease of the amplitude of the PLE signal with the
decrease of $S_z$ (inset of fig.~6(a)): a decrease of the
splitting reduces the spin flip rate and consequently the
efficiency of the exciton transfer. Similarly, when
exciting the state $|+1,S_z=+1/2\rangle$, the exciton
preferentially relax to the dark exciton level
$|-2,S_z=+1/2\rangle$ through a spin flip of the hole, at
lower energy than the bright exciton $|-1,S_z=+1/2\rangle$.
Despite the weak oscillator strength of this level, most of
the resonant PL is then observed on this low energy state.

For a given Mn spin state, the amplitude of the resonant
fluorescence depends on an applied external magnetic field.
This is illustrated in Fig.~7 where the PLE spectra
detected on the low energy line when the excitation laser
is tuned around the high energy level are presented for
different magnetic fields in Faraday geometry. The PL
intensity is divided by two for a magnetic field of about
100mT. This reduction could either be explained by a
reduction of the spin flip rate of the exciton or an
increase of the efficiency of the optical pumping of the Mn
spin. The Zeeman energy of the exciton in this weak
magnetic field is not significant compared to the exchange
field with Mn: the dynamics of the exciton coupled with the
Mn is unlikely to be affected by this small energy change.
Thus, the magnetic field dependence is attributed to an
increase of the optical pumping efficiency. Indeed, the
Zeeman splitting of the Mn in an empty dot cancels the
nondiagonal coupling induced by the tetragonal crystal
field or an anisotropic strained distribution at the Mn
atom location \cite{Qazzaz1995,LeGall2009}. It improves the
Mn spin conservation thus accounting for the increase of
the optical pumping efficiency in a weak magnetic field.

\section{Mn spin dynamics and optical pumping mechanism.}

In the absence of carriers, the dominant spin relaxation
mechanism for the Mn is the coupling to acoustic phonons.
The relaxation rate is then proportional to the density of
states of phonons at the energy of the spin transition.
Thus, the larger the energy difference between initial and
final spin states, the larger the spin-flip
rate\cite{khaetskii2001}. For instance, for diluted Mn
spins under large magnetic field, a B$^5$ increase of the
spin relaxation rate has been reported \cite{Strutz1992}.

At zero magnetic field, Mn spin transitions in the ground
state have an energy splittings of the order of the
strained induced magnetic anisotropy D$_0\approx7\mu eV$
\cite{LeGall2009}. In the presence of the exciton, however,
the Mn splitting is of the order of J$_{hMn}$ which is at
least 10 times larger. This should give relaxation rates
five orders of magnitude larger. In QDs with in-plane
biaxial strains, the inequality $\Gamma_{Mn,G} \ll
\Gamma_{Mn,X}$ is then easily verified and the mechanism of
Mn-spin pumping is relatively insensitive to the value of
the Mn spin lifetime $\Gamma_{Mn,G}$.

To model the optical pumping process, we propose a master equation
for the occupation of the eigenstates of a simplified single
Mn-doped Hamiltonian QD that features the projection along the $z$
axis of the spin $1/2$ electron $S^e$, the pseudospin $J=3/2,J_z=\pm
3/2$ of the heavy hole and the spin $5/2 $ of the Mn:

\begin{eqnarray}
{\cal H}= J_{hMn} S^h_{z} S^{Mn}_z + J_{eh} S^h_{z}S^e_{z}
+ J_{eMn} S^e_{z} S^{Mn}_z \nonumber\\
+ D_0  (S^{Mn}_z)^2
\label{hamil}
\end{eqnarray}

The first term corresponds to the hole-Mn coupling, which
is antiferromagnetic ($J_{hMn}>0$). The second to the
electron-hole exchange, which is ferromagnetic, so that the
dark $\pm 2$ excitons lie below the bright $\pm 1$ exciton
doublet. The third term is the electron-Mn exchange. The
fourth term is the single ion anisotropy adequate for a
strained thin layer \cite{Qazzaz1995}. The spin-flip
contributions, present in the $e-Mn$ case and, depending on
the dot shape, in the $e-h$ and $h-Mn$ coupling, have been
studied in detail elsewhere \cite{JFR06} and, in first
approximation, can be neglected to model the optical
pumping process at zero magnetic field. Hamiltonian
(\ref{hamil}) commutes with $S^{Mn}_z$, $S^e_z$ and $S^h_z$
so that the eigenvalues are $E(S^{Mn}_z,S^e_z,S^h_z)= E_X+
J_{hMn} S^h_{z} S^{Mn}_z+ J_{eh} S^h_{z}S^e_{z}  +  J_{eMn}
S^e_{z} S^{Mn}_z $ where $E_X\simeq 2eV$ is the bare
exciton energy. To reproduce typical experimental spectrum
we take $J_{eMn}=0.05meV$, $J_{hMn}=0.2meV$ and
$J_{eh}=-0.5meV$ \cite{Besombes2004}. The corresponding
spectrum is made of 6 doublets for the bright exciton above
the 6 doublets for the dark excitons. The ground state
spectrum is given by $D_0(S^{Mn}_z)^2$, with $D_0$=7$\mu$eV
\cite{Qazzaz1995}.

\begin{figure}[hbt]
\includegraphics[width=3.5in]{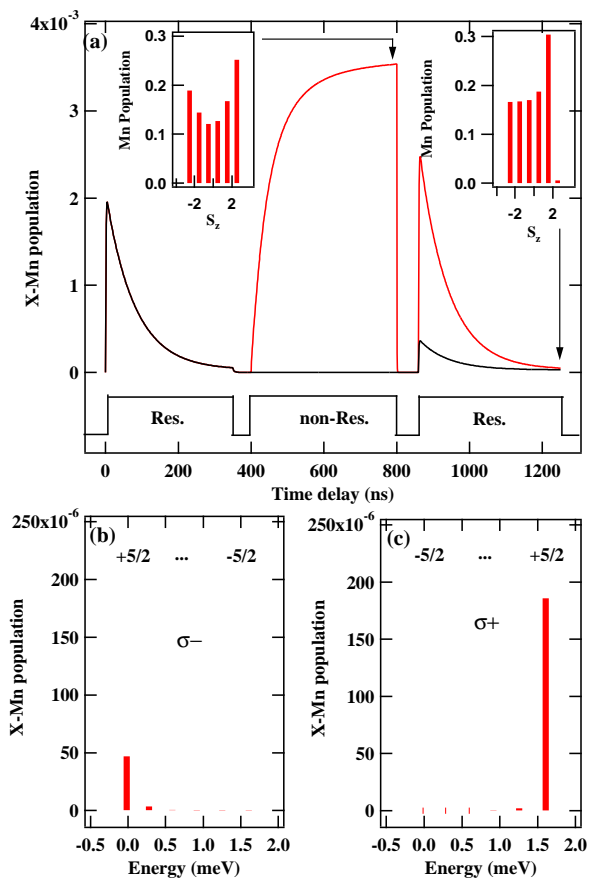}
\caption{(color online) (a) Calculated time resolved PL of
the low energy line of a Mn doped QD under the excitation
sequence used in the resonant pumping experiments (see
Fig.~2). (b) and (c): steady state population of the
$|-1\rangle$ and $|+1\rangle$ excitons respectively at the
end of the resonant pumping pulse on $|+1,S_z=+5/2\rangle$.
The level are displayed as a function of there energy. The
parameters used in the calculation are: $J_{eMn}=-0.05meV$,
$J_{hMn}=0.2meV$, $J_{eh}=-0.5meV$, $D_0=0.007meV$, $T=6K$,
$\Gamma_{e}=0.1ns^{-1}$, $\Gamma_{h}=0.1ns^{-1}$,
$\Gamma_{Mn,X}=0.05ns^{-1}$, $\Gamma_{Mn,G}=0.0001ns^{-1}$,
$\Gamma_{b}=4ns^{-1}$, $\Gamma_{d}=0.1ns^{-1}$,
$\Gamma_{Pump}=0.5ns^{-1}$,
$\Gamma_{Probe}=0.05ns^{-1}$.}\label{fig7}
\end{figure}

The master equation model that we use here is an extension
of the model by Govorov and Kalameitsev (hereafter GK)
\cite{Govorov2005} in their proposal of optical spin
pumping of a single Mn in a QD. In GK, a unique rate is
assigned to transitions between the 24 exciton levels,
complying with principle of detailed balance but neglecting
the dependence of the rates on energy and spin change. Here
we propose a model in which transitions are only permitted
between states that are connected via the flip of a single
spin (the one of either the Mn, the electron or the hole).
This rule is certainly valid in most of spin relaxation
mechanisms and restricts significantly the relaxation
pathways. We still neglect the dependence of the rates on
the energy difference, except for the fact that we use the
principle of detailed balance.  Thus, our model has 4
elementary rates $\Gamma_e$, $\Gamma_h$, $\Gamma_{Mn,X}$
and $\Gamma_{Mn,G}$ for the relaxation of the  spin of the
electron, hole, Mn in the presence of the exciton and Mn
alone. In addition, $S^{Mn}_z$ conserving transitions
between the six ground states and the 12 bright excitons
transitions are described with a laser pumping function
$g(S^{Mn}_z,X)$ where $X=\pm 1$ labels the bright excitons.
We consider that dark excitons are not created under these
excitation conditions and take $g(S^{Mn}_z,X=\pm2)=0$. Mn
spin conserving spontaneous emission from bright $\Gamma_b$
and dark $\Gamma_d$ are also included.

In the weak excitation regime, the PL and optical pumping process
are described by the master equation:

\begin{equation}
\label{rate}
\frac{dP_{n}}{dt}=\sum_{n\prime}(\Gamma_{n\prime\rightarrow
n}P_{n\prime})-(\sum_{n\prime}\Gamma_{n\rightarrow n\prime})P_{n}
\end{equation}

\noindent $n$ and $n\prime$ denoting the 24 exciton-Mn
eigenstates and the 6 Mn spin states. We consider that
these spin-flips are triggered by the emission or
absorption of single acoustic phonons and are then
thermally activated. Here we use $\Gamma_{i,n\rightarrow
n\prime}$=$\Gamma_i$ if
$E_{n,n\prime}=E_{n\prime}-E_{n}\leq0$ and
$\Gamma_{i,n\rightarrow
n\prime}$=$\Gamma_{i}e^{-E_{nn'}/k_BT}$ if
$E_{n,n\prime}>0$ with i=e, h, Mn(X) or Mn and $E_{n}$ the
eigenenergies of the Hamiltonian (\ref{hamil}).

The model permits to consider the two excitation modes used
in the pump-probe experiment: $(i)$ resonant, for which
$g(S^{Mn}_z,X)$ is non-zero for only one of the 12 bright
excitons states and $(ii)$ unpolarized and quasi-resonant
for which $g(S^{Mn}_z,X)$ is non-zero for all the 12 bright
states. The time resolved fluorescence signal detected on
the low energy line of X-Mn calculated using the above rate
equation model is presented in Fig.~8(a). The excitation
conditions used in the calculation presented in fig.~8(a)
reproduce the pump-probe sequence of the time resolved
optical pumping experiments displayed in Fig.~2. Value of
the parameters are listed in the caption of fig.~8. The
exchange integrals are chosen to reproduce typical
exciton-Mn spectra and values for $\Gamma_{Mn,X}$,
$\Gamma_{b}$ and $\Gamma_{d}$ where deduced from time
resolved PL and autocorrelation measurements
\cite{Besombes2008}. A ratio
$\Gamma_{Mn,X}/\Gamma_{Mn,G}\approx500$ is enough to
explain the observed optical pumping efficiency. The
characteristic time of the optical pumping is also well
reproduced with a generation rate $\Gamma_{Pump}$ which
drives the QD close to the saturation.

Fig.~8(b) and 8(c) present the steady state population of
the $|-1\rangle$ and $|+1\rangle$ excitons respectively at
the end of a resonant pumping pulse on
$|+1,S_z=+5/2\rangle$. This population is proportional to
the PL intensity of the bright exciton levels. The
introduction in the model of independent spin-flip rates
for the confined carriers and the Mn qualitatively explain
the observed resonant fluorescence spectra: the scattering
spin-flip processes mainly conserve the Mn spin and an
excitation on $|+1,S^{Mn}_z=+5/2\rangle$ empty the state
$S^{Mn}_z=+5/2$ and induces preferentially PL on
$|+1,S^{Mn}_z=+5/2\rangle$ and $|-1,S^{Mn}_z=+5/2\rangle$.

\section{Conclusion}

In summary, our results demonstrate the spin optical
pumping of a single magnetic atom in a semiconductor host.
Resonant excitation of a Mn-doped QD ground state permits a
high fidelity preparation of the Mn spin. We show that
spin-flip scattered photons can be used to probe the
dynamics of the initialization of the Mn spin during the
optical pumping process. The technique introduced here
allows for an accurate and direct observation of the
exciton spin-flip scattering. These measurements confirms
that, as far as the spin dynamics is concerned, the Mn acts
on the exciton as an effective magnetic field which
increase the probability of single phonon induced
spin-flips. A more efficient Mn spin read-out could be
obtained monitoring directly the photons resonantly
scattered by the optically driven excitonic level.

\begin{acknowledgements}
The authors thanks J. Cibert and D. Ferrand for fruitful
discussions. This work is supported by the French ANR
contract QuAMOS and \emph{Fondation Nanoscience}
(RTRA-Grenoble). JFR acknowledges funding from  MEC-Spain
(Grant Nos. MAT07-67845 and CONSOLIDER CSD2007-0010).
\end{acknowledgements}

\end{document}